\documentclass[aps,prb,reprint,groupedaddress]{revtex4-2}
\usepackage{graphicx}
\usepackage{bm}
\usepackage{xcolor}

\begin{document}

\newcommand{\be}{\begin{equation}}
\newcommand{\ee}[1]{\label{#1}\end{equation}}
\newcommand{\bem}{\begin{eqnarray}}
\newcommand{\eem}[1]{\label{#1}\end{eqnarray}}
\newcommand{\eq}[1]{Eq.~(\ref{#1})}
\newcommand{\Eq}[1]{Equation~(\ref{#1})}
\newcommand{\ua}{\uparrow}
\newcommand{\da}{\downarrow}
\newcommand{\g}{\dagger}
\newcommand{\rc}[1]{\textcolor{red}{#1}}


\title{Ballistic SNS sandwich as a Josephson junction}



\author{Edouard  B. Sonin}
\email[]{sonin@cc.huji.ac.il}

\affiliation{Racah Institute of Physics, Hebrew University of Jerusalem, Givat Ram, Jerusalem 9190401, Israel}


\date{\today}

\begin{abstract}
The paper develops the theory of the ballistic SNS sandwich, in which  the Josephson  effect exists without the proximity effect. The theory  takes into account restrictions imposed by the charge conservation law and the incommensurability of the superconducting gap with the Andreev level energy spacing. This resulted in  revisions of some  conclusions of previous works.  In the one-dimensional case the  Josephson phase of the ground state of the ballistic SNS sandwich is not necessarily zero but may have  any value from  0 to $\pi$. If this value is $\pi$ this is a $\pi$ junction, which was well known before. The  suppression of the supercurrent at  temperatures  on the order or higher than the  Andreev level energy spacing, which was predicted in previous investigations, does not take place in the one-dimensional case.
 
At zero temperature the ballistic SNS sandwich of any dimensionality is not a weak link. This leads to unusual  properties:  the absence of the Josephson plasma mode  localized at the normal layer and the Meisner effect  with the same London penetration depth in the normal and  the superconducting layers.
\end{abstract}


\maketitle

\section{Introduction}

\begin{figure}[!b]
\includegraphics[width=.3\textwidth]{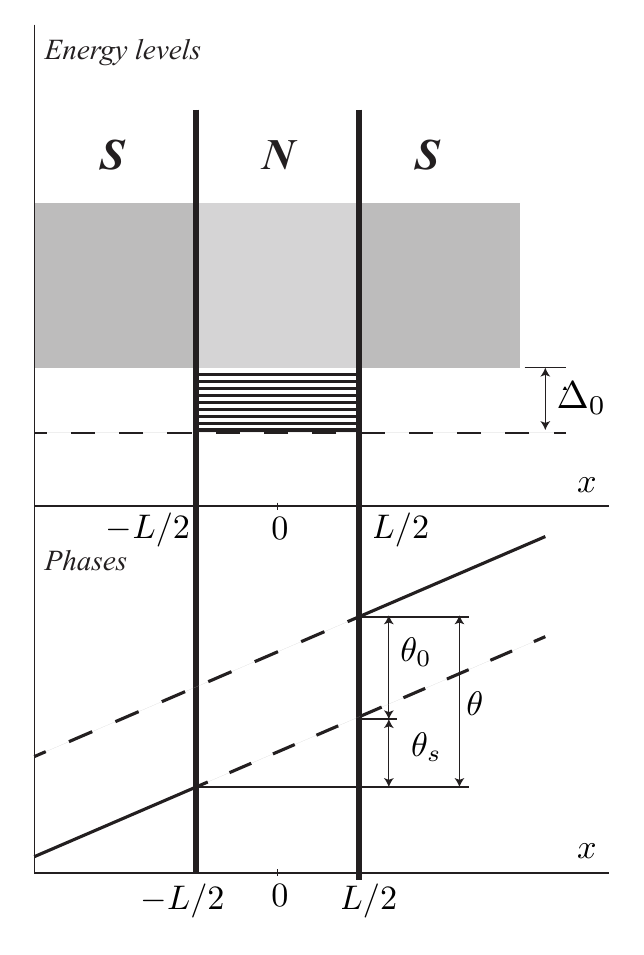}
\caption[]{Energy levels and phases in the SNS sandwich. The interval of continuum states is shaded. The Andreev bound states inside the gap $\Delta_0$ are shown by solid lines. The lower part of the figure shows the bound-state phase $\theta_0$, the superfluid phase $\theta_s$, and the Josephson phase $\theta$.  \label{f1} }
\end{figure}

Originally the Josephson junction was considered as an insulator or normal metal bridge between two superconductors. The Josephson coupling between superconductors was provided due to penetration of the superconducting order parameter into the bridge (proximity effect) if the  bridge is not too long  compared with the coherence length. However, it was noticed long ago  \cite{Kul,Ishi,Bard} that if the bridge is a ballistic  normal metal the Josephson coupling is possible even for rather long bridges.
This was demonstrated in an idealized model of the ballistic SNS sandwich (planar SNS Josephson junction).  There is a normal layer of width $L$ between two superconductors.  The layers are perpendicular to the axis $x$ (Fig.~\ref{f1}). 
The effective masses and Fermi energies are the same in  the superconductors and in the normal
metal. The only difference is that  the pair potential sharply vanishes in the
normal layer  $-L/2<x<L/2$.  Investigations of this model continue up to now \cite{Affleck}.
The ballistic SNS Josephson junction was studied for unconventional pairing in high-$T_c$ superconductors \cite{Ting}.  There were theoretical and experimental investigations  for other materials bridging two superconductors: graphene \cite{Calado,BalGraph}, topological  insulator \cite{Shnir}, and nanotubes \cite{CNT}.

Previous theoretical investigations of the ballistic SNS sandwich  have left some questions unanswered  up to now. \citet{Ishi} noticed that canonical relations for the pair of Hamiltonian conjugated variables ``charge--phase'' were not satisfied. There was a problem with the charge conservation law because the theory postulated some spatial distribution of the order parameter (gap) without solving the  self-consistency equation for gap, which determines this distribution. 
There were also disagreements on the final form of the current--phase relation.

The present paper suggests an approach free of those flaws. In particular,  restrictions imposed by the charge conservation law were checked.   This resulted in a  revision of some previous results.  
The charge conservation law can  be satisfied only taking into account three contributions to the total current: (i) The current induced by the phase gradient in the superconducting layers. We shall call it  the Cooper-pair condensate, or simply  condensate current. (ii) The current, which can flow  in Andreev  states even if the Cooper-pair condensate is at rest  and all Andreev states are empty. It will be called vacuum current.  (iii) The current  induced by nonzero occupation of Andreev states, i.e., by creation of quasiparticles. It will be called excitation current.
The condensate motion produces the same current  in superconducting and normal layers of the SNS sandwich, while  vacuum  and  excitation currents, which are connected with the Andreev states, exists only in the normal layer. The charge conservation law requires that in a stationary state the total current in all layers must be the same.
Thus, the sum  of the  vacuum  and   the excitation currents must always vanish.

Our analysis revealed the effect  of incommensurability of the spectrum gap in the superconducting layers to the Andreev level energy spacing  in one-dimensional (1D) case, when normal and superconducting layers become normal  and superconducting segments of a 1D wire. The effect is important up to high temperatures. Here and later on low  or high temperatures mean temperatures much lower or much higher than the Andreev level energy spacing, but still always much lower than the superconducting gap.  Due to the incommensurability effect,  in the ground state of the SNS sandwich the phase difference $\theta$ across the  SNS sandwich is not necessarily zero, but may vary  from zero  to $\pi$. In the past Josephson junctions  with the ground state at the phase difference $\pm \pi$ were well known and called  $\pi$  junctions.  In analogy with this, we shall call  junctions with the phase difference $\theta$ in the ground state $\theta$ junctions. Josephson $\pi$  junctions were predicted and observed in ferromagnetic junctions \cite{Ryaz},  junctions with unconventional superconductivity \cite{unconv}, quantum dot junctions \cite{Dots}, and SINIS junctions \cite{Volkov}. The transition from 0 to $\pi $ junction was observed in carbon nanotube Josephson junctions \cite{CNT} (see further discussion in the concluding section \ref{Sum}). Another outcome of our analysis is that strong suppression of the supercurrent at temperatures comparable or higher than the  Andreev level energy spacing, which was predicted in previous investigations  \cite{Kul,Ishi,Bard}, does not take place in 1D systems.

Sometimes at currents smaller than critical  values not only the sum of the vacuum and excitation currents vanish, but any of them vanishes separately. This takes place in 1D systems at any temperature and in systems  of any dimensionality at zero temperature. Thus, the charge is transported only by the moving condensate, and the phase distribution does not differ  from the case when the normal layer is replaced by a superconducting layer from the same material as other layers, i.e., does not differ from a uniform superconductor.  
Then the ballistic SNS junction is not a weak link, and therefore, there is no Josephson plasma mode with the frequency much lower than the plasma frequency in the superconducting layer and no suppression of the Meisner effect in the normal layer.   A weak magnetic field penetrates into the normal layer on the same London penetration depth as into the superconducting layers, in contrast to usual Josephson junctions with the Josephson penetration depth much larger than the London penetration depth.  At stronger magnetic fields the Josephson vortices appear with the core size of the order of the normal layer thickness $L$.  Since their energy is lower than the energy of bulk Abrikosov vortices, Josephson vortices are pinned to the normal layer, and the first critical magnetic field for the SNS junction is smaller than that for superconducting bulk, but not so  small as in usual Josephson junctions.

The analysis mostly addresses the 1D case, when only motion along the axis $x$ normal to layers is considered. Its generalization on the 2D and 3D cases is straightforward. Integration over spaces of transverse wave vectors in 2D and 3D cases results in replacement of the 1D electron density by  2D and 3D densities respectively in all expressions for currents, which become current densities.

\section{The Bogolyubov--de Gennes theory}
 \label{BdG}

Since in our model the order parameter $\Delta$ is supposed  to be known we do not need the full BCS Hamiltonian with the  interaction  term quartic in  the electron wave function. It is sufficient to use the quadratic in the wave function second-quantized effective Hamiltonian introduced in the  self-consistent field method  \cite{deGen}. Its density is
\bem
{\cal H}_{eff}={\hbar^2 \over 2m} [\nabla \hat\psi^\g_\gamma(x) \nabla \hat\psi_\gamma(x)- k_f^2 \hat\psi^\g_\gamma(x) \hat\psi_\gamma(x)] 
\nonumber \\
+\Delta \psi^\g_\ua(x) \psi^\g_\da(x)+\Delta^* \psi_\da(x) \psi_\ua(x), 
  \eem{ef}
where $\hat\psi_\gamma^\g(x)$ and $\hat\psi_\gamma(x)$ are operators of creation and annihilation of an electron, and the subscript $\gamma$ has two values corresponding to the spin up ($\ua$) and  down ($\da$).
We address a 1D problem  with the Fermi  wave number $k_f$, assuming that our system is uniform in the plane normal to the axis $x$.  In multidimensional (2D and 3D) systems with the Fermi wave number $k_F$   $k_f =\sqrt{k_F^2 -k_\perp^2}$, where $k_\perp$ is the transverse component of the multidimensional wave vector $\bm k$. 
The complex order parameter, or gap, $\Delta $ can vary in  space.

The quadratic effective Hamiltonian can be diagonalized by the Bogolyubov--Valatin transformation from the free electron operators $\hat\psi_\gamma^\g(x)$ and $\hat\psi_\gamma(x)$ to  the quasiparticle operators $\hat a_{i\gamma}^\g$ and $\hat a_{i\gamma}$:
\bem
\hat \psi_\ua(x) =\sum_{i} \left[u_i(x) \hat a_{i\ua} -v^*_i(x) \hat a^\dagger_{i\da}   \right],
\nonumber \\
\hat \psi_\da(x) =\sum_{i} \left[u_i(x) \hat a_{i\da}+v^*_i(x) \hat a^\dagger_{i\ua}   \right].
    \eem{BV}
For diagonalization of the  effective Hamiltonian the functions $u_i(x)$ and $v_i(x)$ must be stationary solutions of 
the time-dependent  Bogolyubov--de Gennes equations \cite{Kuemmel}:
\bem
i\hbar {\partial u\over \partial  t} ={\delta{\cal H}\over \delta u^*} =-{\hbar^2 \over 2m} \left( \nabla^2 + k_f^2\right) u  + \Delta  v ,
\nonumber \\
i\hbar {\partial v\over \partial  t}  ={\delta{\cal H}\over \delta v^*}={\hbar^2 \over 2m}  \left( \nabla ^2 + k_f^2\right) v  + \Delta^* u .
     \eem{BG}
 The summation over the subscript $i$ means the summation over all bound and continuum states corresponding to stationary solutions of the Bogolyubov--de Gennes equations \eq{BG}.
The Bogolyubov--de Gennes equations
are the  Hamilton equations with  the Hamiltonian (per unit volume) 
\bem
{\cal H}_{BG}={\hbar^2 \over 2m} ( |\nabla u|^2 - k_f^2  |u|^2)-{\hbar^2 \over 2m} ( |\nabla v|^2 - k_f^2  |v|^2)
\nonumber \\
+\Delta u^* v+\Delta^*v^* u.
   \eem{ham}
After the diagonalization the effective  Hamiltonian becomes
\be
{\cal H}_{eff}=\sum_i\varepsilon_i(a^\g_{i\ua} a_{i\ua}+a^\g_{i\da} a_{i\da}-2|v|^2),
  \ee{efd}
where $\varepsilon_i$ is the energy of the $i$th quasiparticle state.

In general the functions $ u(x,t)$ and $v(x,t)$ can be considered as two components of  a spinor wave function,
\begin{equation}
\psi(x,t) = \left( \begin{array}{c} u(x,t)\\  v(x,t) \end{array}
\right), 
     \label{spinor} \end{equation}
describing a state of a quasiparticle, which is a superposition of a state with one particle (upper component~$u$) and a state with
one antiparticle, or hole (lower component~$v$).  The number of particles (charge) is not a quantum number of the state.

The Hamiltonians \eq{ef}  and \eq{ham} are not gauge-invariant, and therefore the total number of electrons (charge) is not a conserved quantity.  Any $i$th solution  of the Bogolyubov--de Gennes equations \eq{BG} satisfies the continuity equation
\be
{\partial n_i\over \partial t} +{1\over e}\nabla j_i= {2i \over \hbar} (\Delta^* v_i^*u_i -\Delta v_iu_i^*),
     \ee{}
where
\be
n_i= |u_i|^2-|v_i|^2 
   \ee{}
is the electron density and
\be
j_i= -{ie\hbar\over 2m} (u_i^*\nabla u_i-u_i\nabla u_i^*) -{ie\hbar\over 2m} (v_i^*\nabla v_i-v_i\nabla v_i^*) 
 \ee{} 
is the electric current. 

Although in the Bogolyubov--de Gennes theory the charge is not conserved, there is another important conservation law for the total probability to find a quasiparticle in the $i$th state somewhere  in the space.  The corresponding continuity equation is
\be
{\partial {\cal N}_i\over \partial t} +\nabla g_i= 0,
     \ee{N-g}
where
\be
{\cal N}_i= |u_i|^2+|v_i|^2 
   \ee{N}
is the quasiparticle density and
\be
g_i= -{i\hbar\over 2 m} (u_i^*\nabla u_i-u\nabla u_i^*) +{ie\hbar\over 2 m} (v_i^*\nabla v_i-v_i\nabla v_i^*) 
 \ee{g} 
is the current, which will be called the quasiparticle flux. While the density $n_i$  is the difference of the densities of particles and holes, the density ${\cal N}_i$ is the sum of these two densities.

The charge conservation law restores  if one solves the Bogolyubov--de Gennes equations \eq{BG} together with the self-consistency  equation. However, we adopt the approach used earlier \cite{Kul,Ishi,Bard}. Instead of solving the self-consistency  equation we simply postulate  the gap $\Delta$ of constant modulus $\Delta_0=|\Delta|$  in the superconducting layers and zero gap inside the normal layer. The model is expected to be valid if the thickness $L$ of the normal layer essentially exceeds the coherence length 
\be
\zeta_0={\hbar v_f\over \Delta_0}. 
    \ee{zeta0}
 
  The total density $n$  and the total charge  current $j$ are  expectation  values for the   operators 
\bem
\hat n(x)=\hat \psi^\dagger_\ua(x)\hat \psi_\ua(x)+\hat \psi^\dagger_\da(x)\hat \psi_\da(x)
 \nonumber \\ 
=\sum_{i}\left[ |u_i(x)|^2  \hat a^\dagger_{i\ua} \hat a_{i\ua}+|v_i(x)|^2  \hat a_{i\da} \hat a^\dagger _{i\da} \right]
\nonumber \\ 
=\sum_{i}\left[ |u_i(x)|^2  \hat a^\dagger_{i\ua} \hat a_{i\ua}-|v_i(x)|^2 \hat a^\dagger _{i\da} \hat a_{i\da} +2 |v_i(x)|^2\right],
  \eem{nOp}
\bem
\hat j= -{ie\hbar\over 2m}\sum_{i} \left[(u_i^*\nabla u_i-u_i\nabla u_i^*)\hat a^\dagger_{i\ua} \hat a_{i\ua} 
\right. \nonumber \\ \left.
+ (v_i^*\nabla v_i-v_i\nabla v_i^*) \hat a^\dagger _{i\da} \hat a_{i\da}-2 (v_i^*\nabla v_i-v_i\nabla v_i^*)\right].
 \eem{n-j} 

There are two additive contributions to the  density, the energy, and the current [Eqs.~(\ref{efd}),  (\ref{nOp}) and (\ref{n-j}) respectively]. One is the vacuum contribution calculated assuming that all energy levels are not occupied (quasiparticle vacuum). This is given by last terms in  equations,  which do not contain any quasiparticle operator.   The other terms in   the  equations  yield  the  contribution of  excitations  due to possible occupation  of energy levels.

In a resting uniform superconductor with the constant $\Delta_0$ solutions of  the Bogolyubov--de Gennes equations are plane waves
\begin{equation}
\left(  \begin{array}{c}
u_0    \\
  v_0 
\end{array}  \right)e^{i k \cdot x-i\varepsilon_0 t/\hbar},
   \label{pwsol}
\end{equation}
where
\be
u_0=\sqrt{{1\over 2} \left( 1+  {\xi\over \varepsilon_0}\right)} ,~~v_0= \sqrt{{1\over 2} \left( 1-  {\xi\over \varepsilon_0}\right)}.
          \ee{u0v0}
The quasiparticle energy is given by the well known BCS expression 
\begin{equation}
\varepsilon_0=  \sqrt{\xi^2 + \Delta_0^2}.
     \label{SpBCS}
\end{equation}
 Here  $\xi = ({\hbar^2 / 2m})(k^2 - k_f^2)\approx \hbar v_f (k-k_f)$  is the quasiparticle energy in the normal Fermi liquid, and $v_f=\hbar  k_f/m$ is the  Fermi velocity.  
The states with positive and the negative signs of $\xi$ correspond to particle-like and the hole-like  branches of the spectrum respectively. Note that mathematically 
 the Bogolyubov--de Gennes equations have solutions with negative and positive energies $\pm \varepsilon_0$. But only solutions with positive energy $\varepsilon_0$ have the physical meaning  \cite{Tin}. In fact, taking into account solutions with negative energy would be a double-counting since hole-like solutions with positive energy but with $k<k_f$ (negative $\xi$) have represent all states  inside the Fermi surface.

\section{Bound Andreev  and continuum states}
 \label{ASSNS}

\subsection{Andreev bound states}

The spectrum and  the wave function for the present model of the SNS sandwich have been already investigated in previous works, and it is sufficient here to  present the resume of these investigations. In the limit of large Fermi wave numbers $k_f \gg \Delta_0 /\hbar v_f$  the Bogolyubov--de Gennes equations of the second order  in gradients are reduced to the equations of the first order. As a result, the boundary conditions on the interfaces between the normal and superconducting layers require the continuity  of the wave function components $u$ and $v$  but not their gradients. 
The components $u$ and $v$ are superpositions of plane  waves with wave numbers close to either only $+ k_f$, or only $- k_f$. This means that at interfaces between normal and superconducting layers only Andreev reflection is possible, which does not change the quasiparticle momentum essentially, but the quasiparticle group velocity changes its sign. 

Because of Andreev reflection,  there are Andreev bound states with energies $0<\varepsilon_0<\Delta_0$ localized in the normal layer.  The wave functions of these states, which satisfy the Bogolyubov--de Gennes equations and the boundary conditions,  are given by
\begin{equation}
\left(  \begin{array}{c}
u    \\
  v
\end{array}  \right)
=\sqrt{N\over 2}\left(  \begin{array}{c}
  e^{\pm {i   \eta\over  2}  \pm {i m\varepsilon_0    \over \hbar^2 k_f}(x-L/2) }   \\    e^{-i\theta_+ \mp {i   \eta\over  2}  \mp {i m\varepsilon_0    \over \hbar^2 k_f}(x-L/2) }
\end{array}  \right) e^{\pm i k_f\cdot x}
  \label{norm}
\end{equation}
 inside the normal layer $-L/2<x<L/2$,
 \begin{equation}
\left(  \begin{array}{c}
u    \\
  v
\end{array}  \right)=\sqrt{N\over 2}\left(  \begin{array}{c}
 e^{\pm {i   \eta\over  2} }    \\
  e^{-i\theta_+ \mp{i   \eta\over  2} } 
\end{array}  \right) e^{\pm i k_f x-(x-L/2) /\zeta}
          \label{sup2}
\end{equation}
 inside the superconducting layer at $x>L/2$,  and
  \begin{equation}
\left(  \begin{array}{c}
u    \\
  v
\end{array}  \right)=\sqrt{N\over 2}\left(  \begin{array}{c}
e^{\mp {i   \eta\over  2} }   \\
 e^{-i\theta_- \pm {i   \eta\over  2} } 
\end{array}  \right)
e^{\pm i k_f  x+(x+L/2)/\zeta }
\label{sup1}
\end{equation}
inside the superconducting layer at $x<-L/2$.
  Here
\begin{equation}
e^{i\eta}=  {\varepsilon_0 + i\sqrt{\Delta_0^2-\varepsilon_0^2}\over \Delta_0},~~\cos\eta={\varepsilon_0\over \Delta_0},~~\sin\eta={\sqrt{\Delta_0^2-\varepsilon_0^2}\over \Delta_0},
 \label{eta} \end{equation}
and $\theta_+$ and $\theta_-$ are the constant order parameter phases in the superconducting layers at $x>L/2$ and $x<-L/2$. 
 The upper  and lower signs correspond to the wave number semi-spaces $k>0$  and $k<0$ respectively. The normalization constant
\be
N={1\over L+\zeta}
    \ee{}
takes into account  the penetration of the bound states into the superconducting layers with the penetration depth
\be
\zeta= \zeta_0{\Delta_0\over \sqrt{\Delta_0^2-\varepsilon_0^2}},
  \ee{zeta}
which diverges when  $\varepsilon_0$ approaches to the gap $\Delta_0$.

The boundary conditions are satisfied at the Bohr--Sommerfeld condition,
\begin{equation}
 \varepsilon_0(s,\pm  \theta_0) ={\hbar v_f\over 2L}\left(2\pi  s+2\eta \pm \theta_0\right) ,
 \label{eps0}     \end{equation}
which determines the energies of the Andreev states. Here $\theta_0=\theta_+-\theta_-$ and  $s$ is an arbitrary integer. The notation  $s$ for  integers  will appear further also in other expressions, although its value would be chosen differently.  The two signs before $\theta_0$ correspond to positive  and negative signs of the 1D wave numbers  in the Andreev states.
Further we shall call the phase difference $\theta_0$  across the normal layer the bound-state phase, because it shifts the bound states with respect to the gap.

\Eq{eps0}  is not an expression  but an equation for  $\varepsilon_0$, since $\eta$ depends on $\varepsilon_0$.
At small energy $\varepsilon_0 \ll \Delta_0$, $\eta=\pi/2$, and the spectrum of the bound states is
\be
\varepsilon_0 ={\hbar v_f\over 2L}\left[2\pi  \left(s+{1\over 2}\right) \pm \theta_0\right].
      \ee{SNSsp}

At the energy $\varepsilon_0$ close to $\Delta_0$ ($\Delta_0 -\varepsilon_0 \ll \Delta_0$) one can use the approximation
\be
\eta \approx \sqrt{2(\Delta_0-\varepsilon_0)\over \Delta_0}.
   \ee{}
Then solution of \eq{eps0}  for $\varepsilon_0$ yields
\be
\varepsilon_0= \Delta_0 -{\hbar^2 v_f^2\over 2\Delta_0 L^2}\left\{\sqrt{1+{\Delta_0 L\over \hbar v_f} [ 2\pi (s +\alpha)  \mp \theta_0]}-1\right\}^2,
   \ee{jsd}
where  $\alpha$ is the fractional part of the ratio
\be
{\Delta_0  L\over \pi \hbar v_f}=s +\alpha.
     \ee{inc}
An integer $s$ is chosen so that $0 <\alpha <1 $. The parameter $\alpha$ is the measure of incommensurability of the gap $\Delta_0$ with the level energy spacing.

The charge current in the occupied  $s$th Andreev state is determined by the canonical relation
\be
j_\pm(s)={2e\over \hbar}{\partial  \varepsilon_0(s,\pm  \theta_0)\over \partial \theta_0}=\pm {ev_f\over L+\zeta}.
  \ee{js}
 The factor 2 takes into account that $\theta_0$ is the phase of a Cooper pair but not of a single electron.  
As expected, this expression fully agrees with  the fact  that
the mass current is the total momentum $\hbar k_f$ in the state divided by the size $L+\zeta$ of the bound state.

The existence of charge current in a bound state is the consequence of the absence of the charge conservation law in our model. At the same time, the quasiparticle flux given by \eq{g} vanishes in accordance with the conservation law \eq{N-g} for the total number of quasiparticles.

\subsection{Continuum states}

Delocalized continuum states with $\varepsilon_0 >\Delta_0$  are scattering states. For  a quasiparticle  ($\xi>0$) incident from left and  propagating from $x=-\infty$ to $x=\infty$ the wave  function is
\bem
\left(  \begin{array}{c}
u_0(\xi)   \\
 v_0(\xi) e^{-i\theta_-}
\end{array}  \right)e^{i \left(k_f+ {m\xi\over \hbar^2k_f}  \right)x}
 \nonumber \\ 
 +\,r\left(  \begin{array}{c}
u_0(-\xi)    \\
 v_0(-\xi)  e^{-i\theta_-} 
\end{array}  \right)e^{i \left(k_f- {m\xi\over \hbar^2k_f}  \right)x}
      \eem{}
for $x<-L/2$, and
\be
t\left(  \begin{array}{c}
u_0(\xi)   \\
 v_0(\xi) e^{-i\theta_+}
\end{array}  \right)
e^{i \left(k_f+ {m\xi\over \hbar^2k_f}  \right)x}
      \ee{}
for $x>L/2$. Here $t$ and $r$ are amplitudes of transmission and
reflection determined from the continuity
of spinor components at $x=\pm L/2$  \cite{Bard}. As in the case
of  bound states, the analysis considers only the Andreev reflection.
The reflection and the transmission probabilities are
\begin{equation}
R(\theta_0)=|r|^2= \frac{\Delta_0^2\left[1-\cos\left({2\varepsilon_0m  L\over \hbar^2 k_f}-\theta_0\right)\right]}{2\varepsilon_0^2 -\Delta_0^2-\Delta_0^2\cos\left({2\varepsilon_0m  L\over \hbar^2 k_f}-\theta_0\right)} ,
\label{ref} \end{equation}
\begin{equation}
{\cal T}(\theta_0)=|t|^2= \frac{2(\varepsilon_0^2-\Delta_0^2)} {2\varepsilon_0^2 -\Delta_0^2-\Delta_0^2\cos\left({2\varepsilon_0m  L\over \hbar^2 k_f}-\theta_0\right)}.
\label{T} \end{equation}
 The spinor  in the normal layer $-L/2<z<L/2$ is given by the same expression as \eq{norm} for the bound state, but with different normalization constant $N={\cal T}$. 

Similar expressions with the same  $R(\theta_0)$ and ${\cal T}(\theta_0)$   can be derived for a quasihole ($\xi<0$) incident from right and moving to left. For a quasiparticle incident from right and a hole incident from left   the reflection and the  transmission probabilities are $R(-\theta_0)$ and ${\cal T}(-\theta_0)$.

The transmission probability differs from unity in the energy
interval of the order $\Delta_0$ small with respect to the Fermi energy $\varepsilon_f=\hbar^2 k_f^2/2m$. The condition $R+{\cal T}=1$ follows from the  conservation law for the number of quasiparticles, which leads to the constant quasiparticle flux $g$ in the whole space [see \eq{g}]. The scattering delocalized states in the SNS sandwich were determined for $\theta_0=0$ by  \citet{Bard} and for $\theta_0\neq 0$ in Refs.~\onlinecite{Son13,EBS}.

One can transform expressions for $R$ and ${\cal T}$ demonstrating their  dependence on the incommensurability parameter $\alpha$ introduced in \eq{inc}:
\begin{equation}
{\cal T}= \frac{2(\varepsilon_0^2-\Delta_0^2)} {2\varepsilon_0^2 -\Delta_0^2-\Delta_0^2\cos\left[{2(\varepsilon_0-\Delta_0)m  L\over \hbar^2 k_f}+2\pi \alpha-\theta_0\right]}.
     \label{Ta} \end{equation}
The reflection  probability can be transformed similarly. The both probabilities rapidly oscillate  as functions of the energy, and at large $L$ one may average over these oscillations neglecting variation of the energy $\varepsilon_0$ within the short oscillation period. The averaged reflection  probability is
\bem
\bar {\cal T} ={1\over 2\pi} \int\limits_{-\pi}^\pi\frac{2(\varepsilon_0^2-\Delta_0^2)d\phi} {2\varepsilon_0^2 -\Delta_0^2-\Delta_0^2\cos\phi}
\nonumber \\
= \frac{2(\varepsilon_0^2-\Delta_0^2)} {\sqrt{(2\varepsilon_0^2 -\Delta_0^2)^2-\Delta_0^4}}={\sqrt{\varepsilon_0^2-\Delta_0^2}\over \varepsilon_0}.~~
     \eem{barT}
After averaging  neither the incommensurability parameter $\alpha$, nor the phase $\theta_0$ influence contributions of continuum states to the transport process.

\section{The ground state (quasiparticle vacuum)}

\subsection{Vacuum current}

\begin{figure}[!t]
\includegraphics[width=0.4
\textwidth]{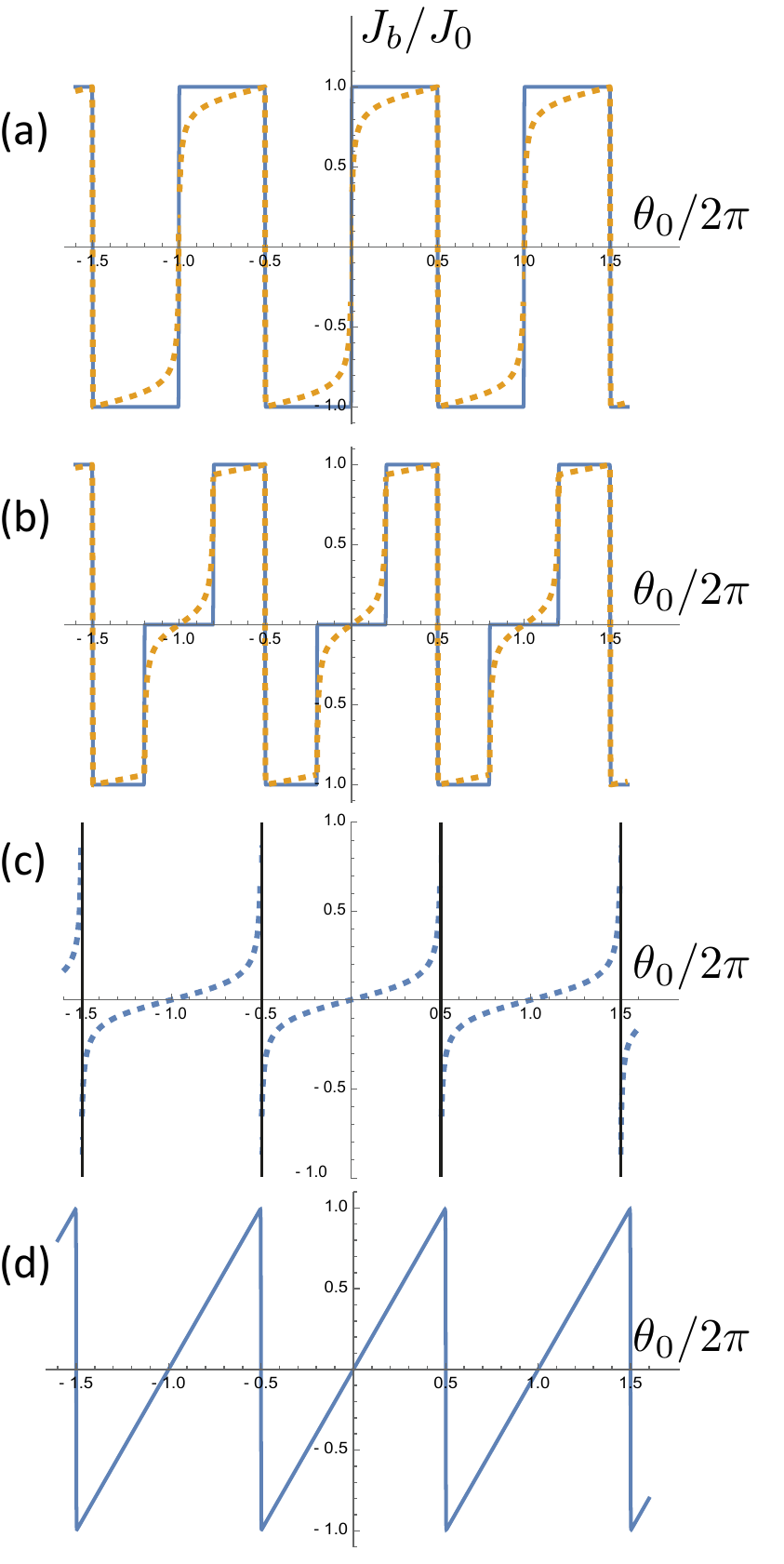} 
 \caption{The vacuum current and energy  vs. the bound-state phase $\theta_0$.  Currents calculated  neglecting or taking into account penetration of Andreev states into superconducting layers at $L/\zeta_0=50$ are shown by solid and dashed lines respectively.  The plots for $\alpha$ and $1-\alpha$ are identical.  (a) $\alpha=0$. (b) $\alpha=0.2$.  (c) The current and the energy averaged over $\alpha$. \label{f2}}
 \end{figure}

In the ground state  in the superconducting layers  the electron fluid is at rest, and there are no currents.  Mathematically in our model the bound-state phase $\theta_0$  is an independent parameter, and the energy  of Andreev states  depends  on it. In order to determine the ground state, one should find the $\theta_0$-dependent energy 
of Andreev states and minimize it with respect  to $\theta_0$.  Since the charge current is determined by the derivative of the energy with respect  to $\theta_0$, after minimization the current vanishes as it should be  in the ground state.

Further we consider the vacuum not in the ground state when the current does not vanish. Frequently the terms ``vacuum'' and ``ground state'' are considered as synonyms. But we define vacuum as a broader term meaning the quasiparticle vacuum when all Andreev levels are empty.

Neglecting the penetration depth $\zeta$ in \eq{js}, the total current of all bound states vanishes if the numbers of states with positive and negative momenta [two signs in \eq{js}]  are equal (the sum of the numbers of states is even),  and they cancel one another.  This is the case at phases $\theta_0=0$ and $\theta_0=\pm\pi$. However, at tuning the phase $\theta_0$ energy levels move. At the both edges of the  Andreev energy spectrum  $\varepsilon_0=0$ and $\varepsilon_0=\Delta_0$  some levels can exit from the gap and some  new levels can enter it. If $\alpha=1/2$ the entrance and the exit processes at the two edges are synchronized: at $\theta_0 =\pm \pi$  a level exits (enters) at the lower edge $\varepsilon_0=0$ [see  \eq{SNSsp}]
 and simultaneously a level enters (exits) at the upper edge $\varepsilon_0=\Delta_0$ [see  \eq{jsd}]. The numbers of states with positive and negative momenta remain equal, and the total current  vanishes.  At $\alpha \neq 1/2$ levels enter or exit at the lower edge of the Andreev spectrum at $\theta_0 =\pm \pi$ as before, but levels cross the upper edge at  $\theta_0 =\pm 2\pi \alpha$.  At $-\pi<\theta_0 <-2\pi\alpha$ and 
 $\pi>\theta_0 >2\pi\alpha$ there is one state with a positive or negative momentum without its counterpart with an opposite-sign momentum. This means that the total momentum is $\pm\hbar k_f$ and the total electric current is $\pm ev_f/L$. Eventually the total vacuum current is
\bem
J_v=-\sum_s[j_+(s)+j_-(s)]= J_0\sum_s\left\{\mbox{H}[\theta-2\pi( s+\alpha)] 
\right. \nonumber \\ \left.
+\mbox{H}[\theta-2\pi (s+1-\alpha)]- 2\mbox{H}\left(\theta-2\pi s-\pi\right)\right\},~~~
  \eem{Jb}
where  $\mbox{H}(q)$ is the Heaviside step function and
\be
J_0= {ev_f\over L}= {\pi e\hbar n_0\over 2mL}.
      \ee{kfn}
Deriving \eq{Jb} we  took into account that  at any $s$  and sign of $\theta$ there are two  states corresponding to two spin values and that according to \eq{n-j} the vacuum current at an  Andreev state is 
two times less and has an opposite sign than the quasiparticle current $j_\pm(s)$. The factors 2 and 1/2 cancel one another.

In \eq{kfn} the relation $k_f=\pi n_0/2$ between $k_f$ and the 1D electron density $n_0$ was  used. After this substitution the formula becomes valid also for 2D and 3D systems bearing in mind that at this generalization $n_0$ and $J_v$ become the electron density and current density in 2D and 3D systems respectively.

The stepwise dependence  of the current $J_v$ on the phase $\theta_0$ at various $\alpha$ is shown in Figs.~\ref{f2}(a)--(d) by solid lines. At $\alpha=1/2$ when the Andreev  levels cross the  lower  ($\varepsilon_0 =0$) and the upper ($\varepsilon_0 =0$) edge of  the gap synchronically the vacuum current vanishes  except for the phases $\theta_0 =2\pi\left(s+{1\over 2}\right) $. At these phases the vacuum current is proportional to the derivative of the $\delta$-function $\delta\left[\theta_0 -2\pi\left(s+{1\over 2}\right)\right]$.

\begin{figure}[!t]
\includegraphics[width=0.4
\textwidth]{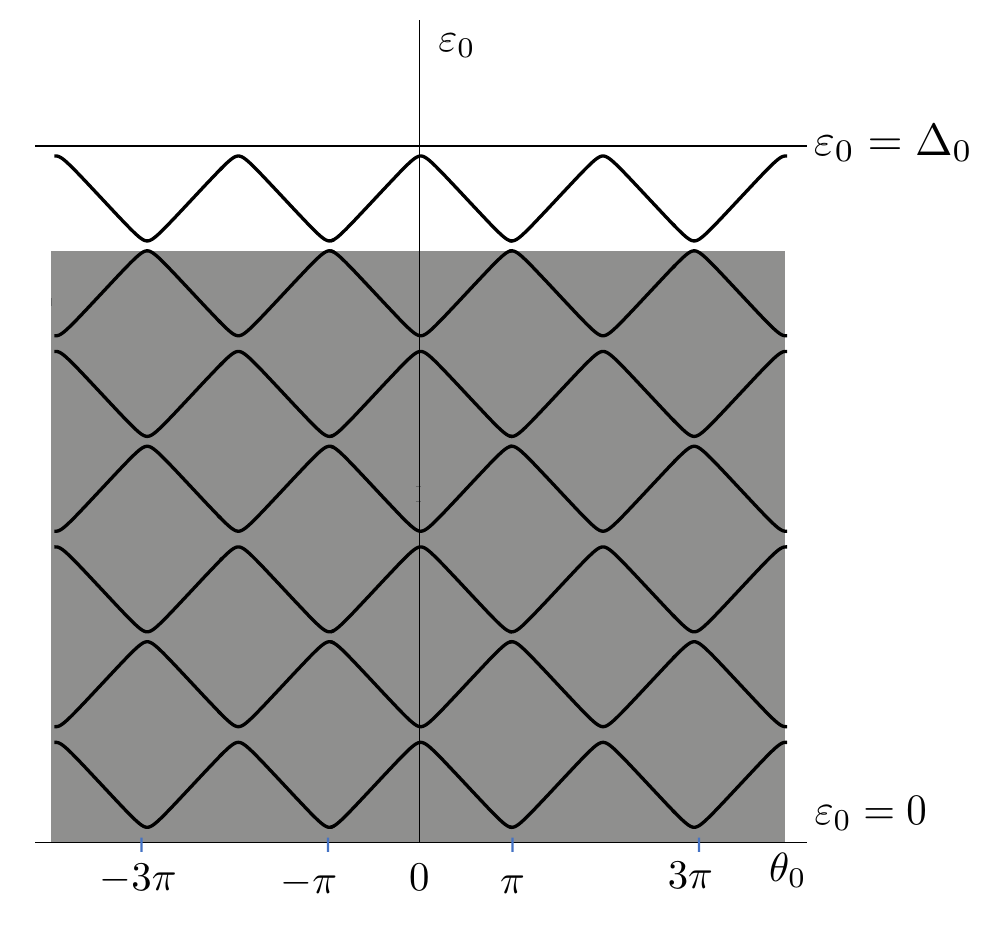} 
 \caption{The Andreev spectrum variation at tuning the bound-state phase $\theta_0$ at  $\alpha =0$.  In the shaded area the number of Andreev states with energies growing  and decreasing with $\theta_0$ are equal and the total current in these states vanishes (see  the text). Only the unshaded band closest to the gap edge $\varepsilon_0=\Delta_0$ is responsible for the total current 
 periodical dependence on  $\theta_0$. \label{f2b}}
 \end{figure}

In our derivation  of  the vacuum current dependence  on $\theta_0$ we used the concept of the spectral  flow, which  is rather popular in the analysis of SNS junctions (see, e.g., Refs. ~\onlinecite{MakhVol,stone96}). The concept assumes that tuning of the phase  $\theta_0$ leads to steady motion of Andreev levels, which cross the whole  gap, i.e., enter the gap on one gap edge and exit from the gap  on the other edge. However, this picture is  valid only in the limit of infinite Fermi wave number when the Andreev level are degenerate at the phases  0 and $\pi$. Even small corrections to this limit lift this degeneracy introducing small gaps   at the phases  0 and $\pi$. As a result,  at phase  tuning the Andreev levels do not cross the gap but oscillate within bands separated by the aforementioned small gaps. Our conclusions remain valid even after this modification of topology of Andreev levels. This illustrated in Fig.~\ref{f2b} for  the case $\alpha =0$ shown in Fig.~\ref{f2}(a). Figure~\ref{f2b} shows the variation of the Andreev-level energies with varying phase $\theta_0$. In  shaded  part  of the spectrum at any phase the numbers of levels with positive and negative slope (i.e., with positive and negative currents) coincide. Thus, contributions of these levels  to the total vacuum current vanish. The  variation of the total vacuum current  with the phase is determined only by the contribution of the unshaded band closest to the gap edge. This  contribution (taking into account that for any Andreev state the vacuum current differs from the current  of the occupied state by the factor -1/2) coincides  with that shown by a solid  line in  Fig.~\ref{f2}(a).

The periodic dependence of the current on the incommensurability parameter is  fragile. In 2D and 3D systems integration over the transverse components of the wave vectors  should  wipe out this dependence.  So, it is reasonable to consider the current  averaged over $\alpha$ in the interval from 0 to 1. After averaging the vacuum current in the interval $-\pi <\theta_0 <\pi $ is
\be
J_v= J_0{\theta_0\over \pi}.
      \ee{JbAv}
The periodical saw-tooth dependence of the current  $J_v$ on $\theta_0$ is shown  in Fig.~\ref{f2}(c).

However, the penetration depth $\zeta$ diverges at $\varepsilon_0 \to \Delta_0$. According to    \eq{js},   at $\zeta \to \infty$ the current in the bound state crossing the upper gap edge vanishes. 
Therefore, we performed a more accurate calculation  in this limit.  At $\Delta_0 -\varepsilon_0 \ll \Delta_0$ the spectrum of bound state is described by \eq{jsd}, and the total current in all bound states is
\begin{widetext}
\bem
J_v= -{J_0\over 2}\left\{ \left[1- \frac{1 }{\sqrt{1+{\Delta_0 L\over \hbar v_f} ( 2\pi \alpha  - \theta_0)}}\right]\mbox{H}( 2\pi \alpha  - \theta_0)
-\left[1- \frac{1 }{\sqrt{1+{\Delta_0 L\over \hbar v_f} ( 2\pi \alpha  + \theta_0)}}\right]\mbox{H}( 2\pi \alpha + \theta_0)
\right. \nonumber \\ \left.
+\zeta\left({1\over 2}, {\hbar v_f \over 2\pi\Delta_0 L}+1+\alpha -{\theta_0\over 2\pi}  \right)-\zeta\left({1\over 2}, {\hbar v_f \over 2\pi\Delta_0 L}+1+\alpha +{\theta_0\over 2\pi}  \right)\right\}.
  \eem{}
\end{widetext}
Here
\be
\zeta\left(z,q\right)=\sum_{s=0} ^\infty{1\over (q+s)^z}
      \ee{}
is Riemann's zeta function \cite{5}. The series for Riemann's zeta function at $z=1/2$ diverges, but the series for a difference of zeta functions with different arguments $q$  converges at large  $s$, which, nevertheless, correspond to energies satisfying the condition $\Delta_0 -\varepsilon_0 \ll \Delta_0$. Therefore,  one can use the infinite series with $s\to \infty$.  The vacuum current $J_v$ calculated taking into account penetration of Andreev states into superconducting layers at $L/\zeta_0=50$ is shown in Fig.~\ref{f2}(a)--(c)  by dashed lines. Summarizing, the divergence of the penetration depth at $\varepsilon_0 \to \Delta_0$ smears the current jump at crossing of the gap edge $\varepsilon_0 =\Delta_0$ by the Andreev level transforming it into a smooth  crossover. But the width of the crossover is small compared to the distance between levels and can be ignored in the limit  $L \to \infty$.

\subsection{Vacuum density}

In the ballistic regime the boundary conditions on the interface affect the wave function in the whole bulk, but it is natural to expect that the average density  in the vacuum  in the ballistic  and the diffusive regime do not differ and are fully determined by the volume of the Fermi sphere  as Luttinger's theorem \cite{Lutting} states.
This also follows from the principle that although dissipative processes are necessary for relaxation to the ground state, the final ground state itself  is not determined by these processes.  Nevertheless, it is useful to check this principle for the SNS sandwich, although this is a check  of our analysis rather than of the principle itself. 

In the superconducting layers at $x<-L/2$ and $x>L/2$  all states are delocalized and form the continuum. For the determination of the vacuum particle density one  can replace in \eq{nOp}  summation  by integration, and the total vacuum density for two spins  and all possible directions of motion of incident quasiparticles and quasiholes  is 
\be
n_0= {1\over \pi}\int_{-\infty}^\infty |v|^2 dk= {1\over \pi \hbar v_f}\int_{-\infty}^\infty |v|^2 d\xi,
      \ee{} 
where 
\bem
|v|^2 ={|v_0|^2\over 4}\left[2+R(\theta_0)+{\cal T}(\theta_0)
\right. \nonumber \\ \left.
+R(-\theta_0)+{\cal T}(-\theta_0)\right]=|v_0|^2={1\over 2}\left(1-  {\xi\over \varepsilon_0}\right).
     \eem{}
 The value of $n_0$ coincides with the density  $n_0 =2k_f/\pi$ in a uniform superconductor. So,  scattering does not affect the average density  $n_0$ in the superconducting layers.

We start the estimation of the density in the normal layer $-L/2<x<L/2$  from the contribution of the Andreev bound states. Any bound state is a superposition of a particle state and of a hole state with equal probability 1/2. Thus, in the normal layer the contribution of Andreev states to the vacuum density is simply a half of  the number of bound states per unit length:
\be
n_{0b} = {2k_f\over \pi} {\Delta_0\over \varepsilon_f}=n_0 {\Delta_0\over \varepsilon_f}.
      \ee{nBS}

The contribution of the continuum states in the normal  layer is
\be
n_{0c}=  {1\over \pi \hbar v_f}\int_{-\infty}^\infty |v|^2 \bar {\cal T}d\xi=n_0\left(1 - {\Delta_0\over\varepsilon_f}\right).
     \ee{Tav}
The averaged transmission  probability $\bar {\cal T}$ is given by \eq{barT}. Together  with the contribution  \eq{nBS} the total density  $n_0=n_{0b}+n_{0c}$ is the same as in uniform normal metals or superconductors with the Fermi energy $\varepsilon_f$. 

\section{Moving Cooper pair condensate}

\subsection{Effect of the Cooper pair condensate motion on  Andreev states (Doppler shift)} 

Let us consider the case of the moving Cooper pair condensate when in  the superconducting layers there is  an order parameter phase gradient $\nabla\varphi$, which determines the superfluid velocity $v_s$:
\be
v_s ={\hbar \over 2m}\nabla \varphi.
    \ee{} 

\begin{figure}[!b]
\includegraphics[width=0.35
\textwidth]{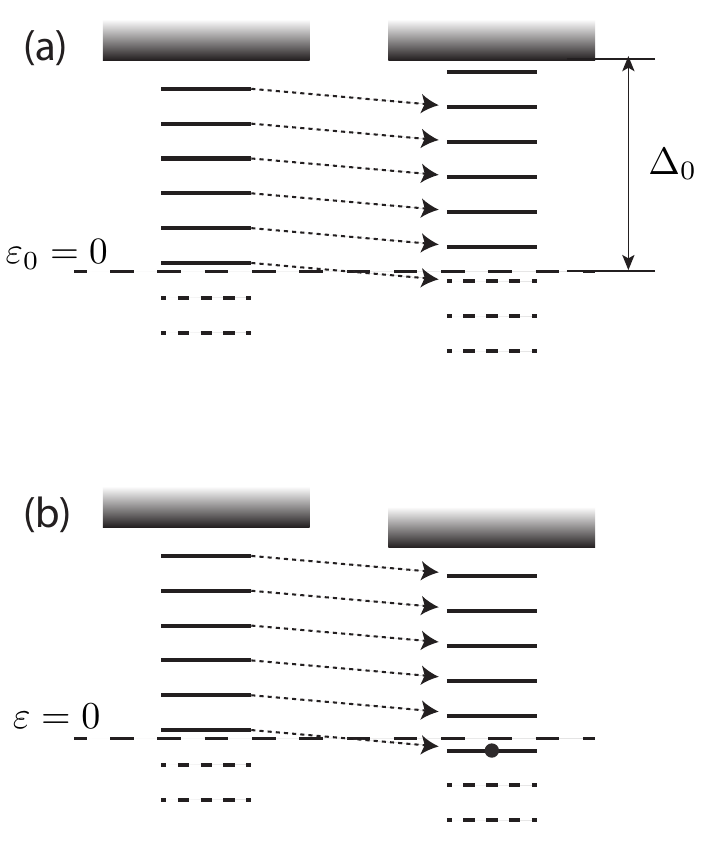} 
 \caption{Tuning of the energies of the Andreev states  by the phases $\theta_0$ and $\theta_s$.   Horizontal  solid lines show unoccupied Andreev levels. A horizontal solid line with a black circle shows an occupied Andreev level. Horizontal dashed lines show ghost levels with negative $\varepsilon_0$, which correspond to mathematically correct solutions of the Bogolyubov--de Gennes equations, but are not considered in the BCS theory as physically real bound states.  Arrowed dashed lines show shifts of levels by tuning the phases $\theta_0$ and $\theta_s$.   (a) Tuning by the phase $\theta_0$  at constant $\theta_s$. The lowest physical level crosses the  energy $\varepsilon_0=0$ and transforms to a ghost level, i.e., disappears.
 (b) Tuning by the phase $\theta_s$ at constant $\theta_0$.  All levels move  together with the gap edge with the energy $\varepsilon_0$ remained constant. The lowest physical level crosses the energy $\varepsilon=0$ and becomes occupied even at zero temperature.
 \label{f2a}}
 \end{figure}

We must solve the Bogolyubov--de Gennes equations \eq{BG} with the  gap
\be
\Delta =\left\{ \begin{array}{cc} \Delta_0e^{i\theta_++ i\nabla \varphi x} & x>L/2 \\  0 & -L/2<x<L/2\\ \Delta_0e^{i\theta_-+ i\nabla \varphi x} & x<-L/2  \end{array} \right. .
   \ee{}
The solution differs from the solution Eqs.~(\ref{norm})--(\ref{sup1}) obtained for the resting condensate by the  presence of the additional factors $e^{imv_s x/\hbar}$ and $e^{-imv_s x/\hbar}$ in the expressions for the components $u$ and $v$ respectively. These factors are cancel in the boundary conditions, and the expressions for $\varepsilon_0$ [Eqs.~(\ref{eps0}), (\ref{SNSsp}) and (\ref{jsd})] and for the reflection and transmission probabilities  [Eqs.~(\ref{ref})--(\ref{Ta})] remains valid. However, the energy $\varepsilon$ of an Andreev state differs from $\varepsilon_0$ by  the Doppler shift:
\be
\varepsilon (s,\pm \theta_0)=\varepsilon_0(s,\pm \theta_0) \pm v_s k_f.
   \ee{}
In particular, at low  energies $\varepsilon_0  \ll \Delta_0$ 
\be
\varepsilon (s,\pm \theta_0)={\hbar v_f\over 2L}\left[2\pi  \left(s+{1\over 2}\right) \pm (\theta_0+\theta_s)\right],
   \ee{eps00}
where
\be
\theta_s={2m  L  v_s \over \hbar}
      \ee{}
is the phase difference across the normal layer as if it were not normal but superconducting (Fig.~\ref{f1}). Therefore, further it will be called superfluid phase.

According to \eq{eps00},  the effects of the bound-state phase $\theta_0$ and the superfluid phase $\theta_s$ on the  energy are additive,  and the energy depends only on their sum. But it  is true  as far as  $\theta_s$   (velocity $v_s$) is small. In general, there is an essential difference between effects of $\theta_0$ and $\theta_s$ on the Andreev spectrum. We saw that variation of $\theta_0$ makes the Andreev levels to move with respect  to the Andreev spectrum edges. As a result, some new levels can emerge and some old ones can disappear. In contrast, variation of $\theta_s$ leads  to the shift of the Andreev spectrum as a whole without changing positions of levels with respect to the Andreev spectrum edges.  This is illustrated in Fig.~\ref{f2a}.
The principle  of the BCS theory that only solutions with positive energies should be taking into account refers to the energy  $\varepsilon_0$, while the Doppler-shifted  energy  $\varepsilon$ can be both positive or negative. If   $\varepsilon$ is negative the level is occupied at zero temperature. This is important for the further analysis.

\subsection{Charge currents due to the motion of the Cooper pair condensate}

The expression for the charge current $J_s$ produced  by the moving Cooper pair condensate follows from \eq{n-j}, in  which only the vacuum contribution is taken into account:
 \bem
J_s= {ie\hbar\over m}\sum_{i}  (v_i^*\nabla v_i-v_i\nabla v_i^*),
 \eem{} 
 where summation is over all bound and continuum states, but the summation over  continuum states can be replaced by integration. Comparing this expression with the vacuum contribution to the electron density in \eq{nOp} one can see that the motion of the Cooper pair condensate produces the $\theta_s$-dependent charge current 
 \be
J_s =e n_0v_s= J_0{\theta_s\over \pi}, 
    \ee{vsJs} 
 in all layers of the sandwich as in a uniform superconductor \cite{Bard}. Thus, the condensate motion induces charge currents satisfying the charge conservation law even in the absence of vacuum and excitation currents in Andreev bound states. This contrasts with the vacuum current, which is produced by the bound-state phase $\theta_0$ only in the normal layer and must be compensates by the excitation current in order to satisfy the charge conservation law.

\section{Excitation contribution to the current }

 Andreev levels in the SNS sandwich are occupied at finite temperatures or  even at zero temperature if the energy $\varepsilon$ of some Andreev levels becomes negative due to the Doppler shift. We consider only temperatures much lower than the gap $\Delta_0$. So quasiparticles  in the superconducting layers are absent. But the temperature can be on the order or higher than the energy distance between Andreev levels. The contribution of  excitations (quasiparticles occupying Andreev levels) to the current at the temperature $T$ is 
\bem
J_q=2J_0 \sum_s \left[\frac{\mbox{H}(s+1/2+\theta_0/2\pi)}{e^{\beta (s+1/2+\theta  /2\pi)}+1}
\right. \nonumber \\ \left.
-\frac{\mbox{H}(s+1/2-\theta_0/2\pi)}{e^{\beta (s+1/2-\theta /2\pi)}+1}\right],
       \eem{temp}
where $\theta=\theta_s+\theta_0$ and 
\be
\beta={\pi\hbar v_f\over LT}.
      \ee{}
The Heaviside functions in numerators provide that only states of the Andreev spectrum with $\varepsilon_0>0$  contribute to the current.
At zero temperature ($\beta\to\infty$) the Fermi distribution function also becomes the Heaviside function,  and the excitation current  is
\be
J_q= -{\theta_s\over |\theta_s|} 2J_0
    \ee{}
in the interval
\be
2\pi\left(s+{1\over 2}\right) -\theta_s <\theta_0<2\pi\left(s+{1\over 2}\right).
    \ee{}
 At high temperatures ($\beta\to 0$) the summation in \eq{temp} can be replaced by integration. In the  interval $|\theta_s|,|\theta_0| <\pi$:
 \be
J_q\approx 2J_0 \int\limits_0^\infty \left[\frac{ds}{e^{\beta (s+\theta  /2\pi)}+1}
-\frac{ds}{e^{\beta (s-\theta /2\pi)}+1}\right]
 =-J_0{ \theta \over  \pi}. 
       \ee{JqH}

\begin{figure}[!b]
\includegraphics[width=0.33
\textwidth]{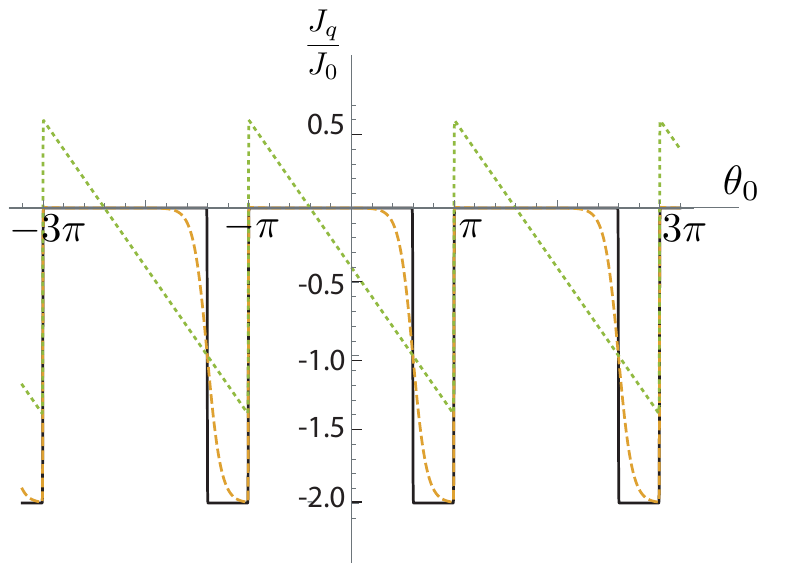} 
 \caption{The excitation  current  vs. the bound-state phase $\theta_0$ at $\theta_s=0.4\pi$.  Solid, dashed, and dotted lines show the current at
zero temperature ($\beta \to \infty$),  low temperature ($\beta=30$), and high temperature ($\beta \to 0$) respectively.  
 \label{f3}}
 \end{figure}

The contribution of quasiparticles at occupied Andreev states to the current is shown in Fig.~\ref{f3}  for $\beta\to \infty$ (zero temperature), $\beta=30$ (low temperature), and $\beta\to 0$ (high temperature) by the solid, dashed, and dotted line respectively.

\section{Charge conservation law and current--phase relation} \label{CurPh}

\begin{figure}
\includegraphics[width=0.45
\textwidth]{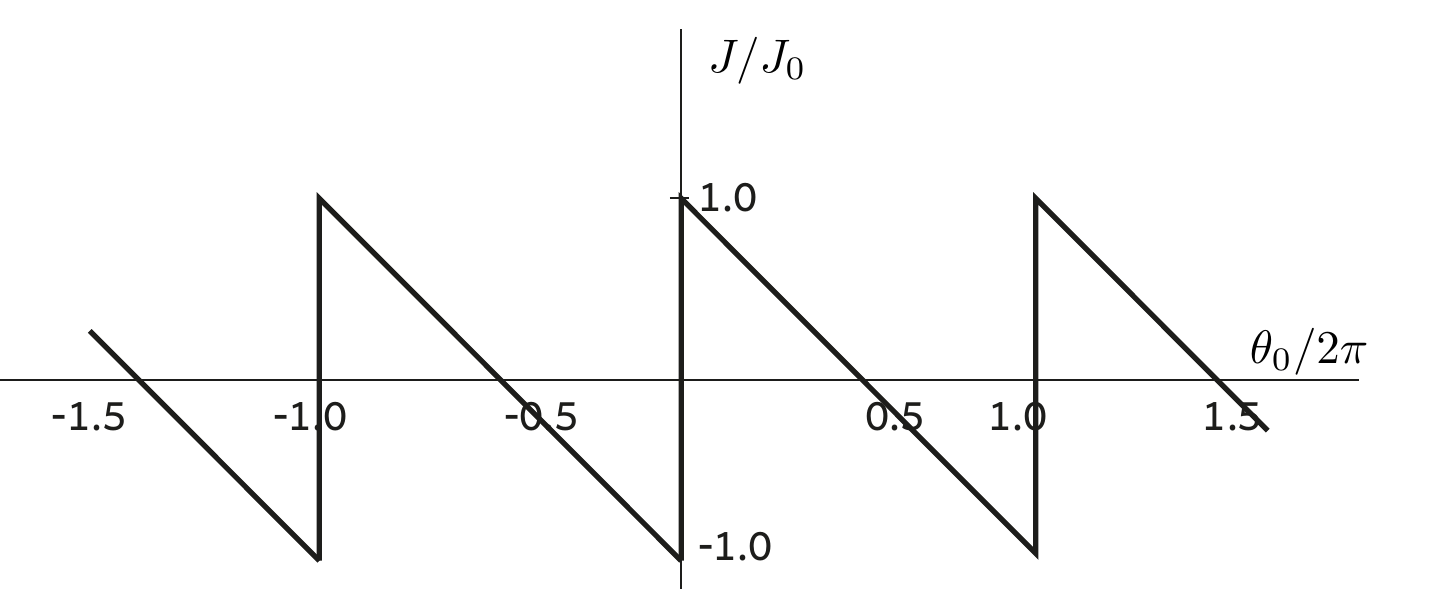} 
 \caption{The current--phase $\theta_0$ relation for zero   temperature.
 \label{f4}}
 \end{figure}

As already mentioned, our model does not satisfy the charge  conservation law, and stationary solutions of the model with different currents in different layers are mathematically correct. However, only solutions, which {\em do satisfy} the charge conservation law, have a physical meaning and  must be chosen. One can meet this requirement by imposing the condition that in the stationary case the current in the normal  layer does not differ from the current in the superconducting layers.  Since the motion of  the condensate with the velocity $v_s$ produces the same current $J_s$ in all  layers, the vacuum and excitation currents $J_v$ and $J_q$ in Andreev states must cancel one another: $J_v+J_q=0$.  So, the total current  $J
=J_v+J_q+J_s$  cannot differ from   $J_s$.

Figure~\ref{f4} shows the current--phase $\theta_0$ relation obtained from  the condition  $J_v+J_q=0$ at zero temperature. It is remarkable that  at zero temperature the current--phase curve $J(\theta_0)$ does not depend on the incommensurability parameter $\alpha$. Along vertical segments of the curve  at  $\theta_0=2\pi s$ both $J_v$ and $J_q$ vanish and all Andreev levels are unoccupied. Compensation of a nonzero vacuum current at $\theta_0 \neq 2\pi s$ by  an excitation current is possible if the lowest-energy Andreev level reaches zero and is at least partially occupied. According to \eq{SNSsp}, this  takes place if $\theta_0+\theta_s=2\pi \left(s+{1\over 2}\right)$. Using \eq{vsJs} one obtains the current 
\be
J=J_s ={J_0\over \pi} \left(2\pi  s+\pi -\theta_0\right).
    \ee{}
at the sloped  segments of the current--phase curve in Fig.~\ref{f4}, which does not depend on the incommensurability parameter $\alpha$. Independence from  $\alpha$ makes averaging over $\alpha$ in 2D and 3D systems unnecessary.  Thus, the current--phase curve shown in Fig,~\ref{f4} is valid for a system of any dimensionality at zero temperature.

\begin{figure}
\includegraphics[width=0.4
\textwidth]{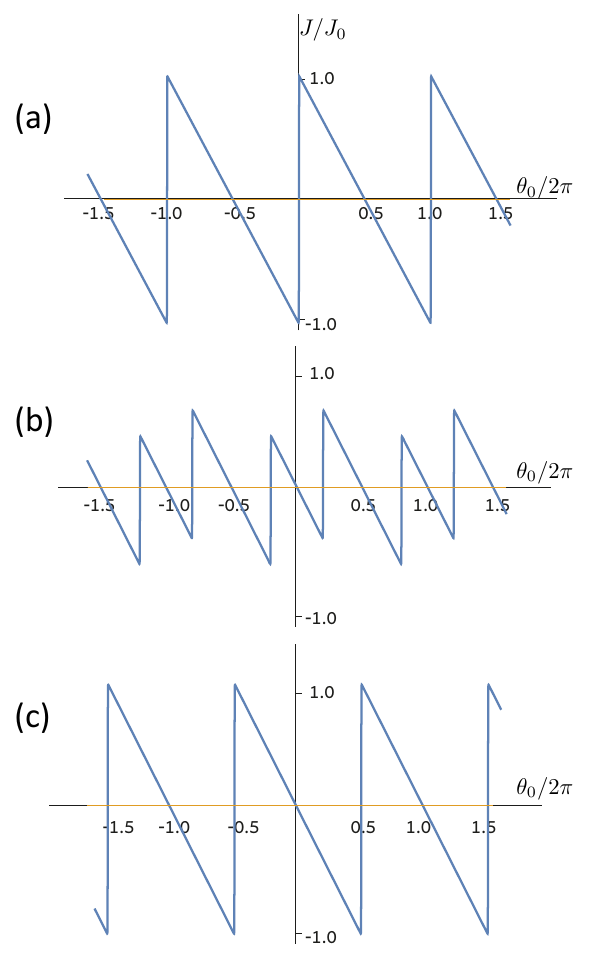} 
 \caption{The current--phase $\theta_0$ relation for high temperature. (a) $\alpha=0$. (b) $\alpha =0.2$. (c) $\alpha=1/2$. 
 \label{f5}}
 \end{figure}

But  at  finite temperature  the current--phase relation {\em does depend} on  $\alpha$. The current--phase curve $J_v(\theta_0)$ at high temperature is shown in Fig.~\ref{f5}  for $\alpha=0$, $\alpha =0.2$, and $\alpha=1/2$. For  $\alpha=0$ the current--phase curve $J_v(\theta_0)$  at high temperature  does not differ from that at zero temperature shown in Fig.~\ref{f4}.

The $\alpha$-dependent current--phase curves in Fig.~\ref{f5} are valid only in the 1D case. After averaging over $\alpha$ in the multidimensional (2D and 3D) systems the vacuum current $J_v$ given by \eq{JbAv} can compensate  the excitation current $J_q$ [\eq{JqH}] only at $\theta_s=0$. So, the supercurrent vanishes in the limit of high temperature when summation in the expression \eq{temp} can be replaced  by integration. At temperature not high enough for validation of this approximation the supercurrent does not vanish completely but  strongly decreases with temperature.

However, the bound-state phase $\theta_0$ is not a phase, which must be used in the canonical description of the  Josephson junction by the pair of conjugate variables ``charge--phase''. The proper phase is the total phase difference across the normal layer $\theta=\theta_0+\theta_s$, which  we call Josephson phase (Fig.~\ref{f1}). The time derivative of the phase $\theta$ determines the voltage drop across the normal layer:
\be
V={\hbar \over 2 e} {d\theta\over dt}.
  \ee{}

\begin{figure}[!t]
\includegraphics[width=0.4
\textwidth]{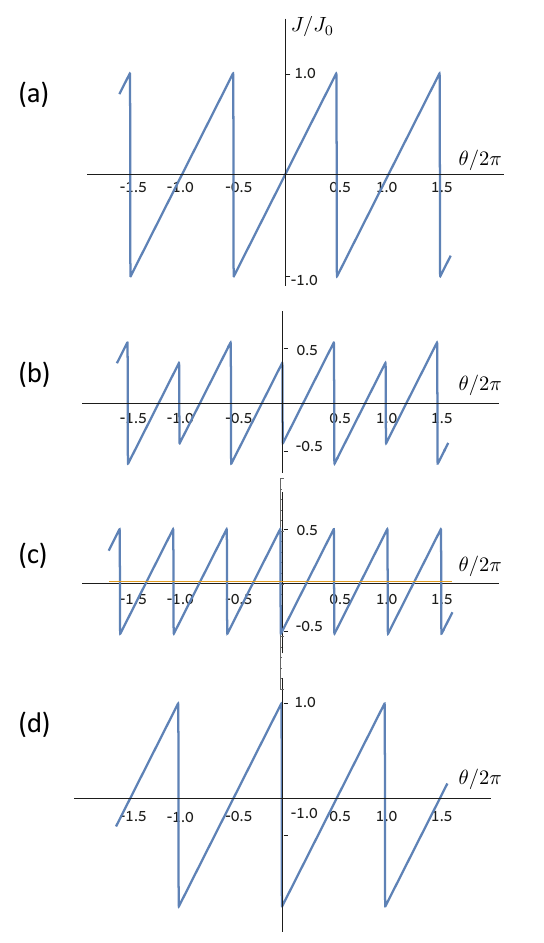} 
 \caption{The current--phase $\theta$ relation for high temperature.  (a) $\alpha=0$. The same curve describes the current--phase relation at  zero temperature, which is independent from $\alpha$. 
 (b) $\alpha =0.2$. (c) $\alpha=0.25$. (d) $\alpha=1/2$.
 \label{f6}}
 \end{figure}

Figure \ref{f6} shows the current--phase relation for the Josephson phase $\theta$  at  various values of $\alpha$ at high temperature. In the phase interval $(-\pi,\pi)$ it is given by
\be
J(\theta)= J_0 \left({\theta\over \pi}-2\alpha {\theta\over |\theta|}\right).         
      \ee{}
The critical Josephson current  (its maximum value) depends on $\alpha$:
\be
J_c= J_0 \times \left\{   \begin{array}{cc} 1- 2\alpha &  \alpha<{1\over 4} \\  2\alpha &  \alpha>{1\over 4} \end{array}.                     \right.
      \ee{crCur}

According to Fig.~\ref{f6}, at any nonzero $\alpha$ the current at small positive (negative) $\theta$ becomes negative (positive). This means that at $\alpha\neq 0$ the Josephson energy 
\be
E_J={\hbar \over 2e}        \int\limits^\theta J(\theta)d\theta, 
      \ee{EJ}
has not a minimum but a maximum. The energy minimum (ground state) is at the phase $\theta =2\pi \alpha$. At varying $\alpha$  from 0  [Fig.~\ref{f6}(a)]  to 1/2  [Fig.~\ref{f6}(d)]   the phase $\theta$ in the ground state varies from 0 to $\pi$. The case $\theta =\pi$ corresponds to a $\pi$ junction well known in the past (see Introduction). In general, one  can call junctions  with the nonzero $\theta$ in the ground state $\theta $ junctions. The current--phase curve of the $\pi/2$ junction ($\alpha=1/4$) in Fig.~\ref{f6}(c)  is periodical with the period $\pi$ instead of $2\pi$ and the  critical current has a minimum, which is two times smaller than that  for 0 and $\pi$ junction ($\alpha$=0 or 1/2).

The current--phase relation shown  in Fig.~\ref{f6}(a), which  is valid for $\alpha=0$ for 1D  systems at high temperature and for any $\alpha$ and any dimensionality at zero temperature, does  not differ from the current--phase relation obtained by \citet{Bard} at zero temperature. Our analysis of multidimensional (2D and 3D) systems also confirms their conclusion that  the  supercurrent vanishes in the limit of temperatures much higher than the Andreev level energy spacing. However, our physical picture of the phenomenon differs  from theirs.  \citet{Bard} took into account  the current $J_s$ produced by the condensate motion and the excitation current $J_q$, but  ignored the vacuum current $J_v$ determined by the phase $\theta_0$ absent in their analysis. The charge conservation law requires that  the sum  $J_v$ and $J_q$  must vanish. The analysis of Ref.~ \onlinecite{Bard}  does not meet this requirement. The difference between the physical pictures is important for 1D systems at high temperatures. In this case  suppression  of the supercurrent at high temperature predicted by \citet{Bard} is not valid.

\section{Some properties  of the SNS sandwich as a Josephson junction}

\subsection{The nonstationary Josephson effect at  current bias}

Let us consider the SNS sandwich shunted by ohmic resistance $R$ at the current bias $I$ exceeding the critical one.  The general expression for the average voltage for the overdamped Josephson junction is  \cite{Tin}
\be
\bar V = \frac{2\pi R}{\int_{-\pi}^\pi {d\theta\over I-J(\theta)}}.
    \ee{}
For the current--phase relations at high temperature shown in Fig.~\ref{f6} this yields the $VI$ curve
\be
\bar V = \frac{2 R J_0}{\ln {(I+2\alpha J_0)[I+(1-2\alpha) J_0]\over (I-2\alpha J_0)[I-(1-2\alpha) J_0]}}.
    \ee{}
Using the expression \eq{crCur} for the critical current one obtains
\be
\bar V= 2RJ_c \times \left\{   \begin{array}{cc} \frac{1}{\ln {\left[(1-2\alpha)I+2\alpha J_c)\right](I+ J_c)\over \left[(1-2\alpha)I-2\alpha J_c)\right](I- J_c)}}&  \alpha<{1\over 4} \\ \\ \frac{1}{\ln {\left[2\alpha I+(1-2\alpha) J_c\right](I+ J_c)\over \left[2\alpha I-(1-2\alpha)  J_c\right](I- J_c)}}&  \alpha>{1\over 4} \end{array}.                     \right.
      \ee{}
We remind that  for $\alpha=0$  the current--phase relations at zero and high temperature do not differ, and the $VI$ curve is
\be
\bar V = \frac{2R J_c}{\ln {I+J_c\over I-J_c }}.
    \ee{}
All  curves follow the Ohm law $ \bar V =RI$ at  $I\gg J_c$. We note for comparison that for Josephson junctions with the sinusoidal  current--phase relation the $VI$ curve is  $\bar V =R\sqrt{I^2-J_c^2}$ \cite{Tin}.

\subsection{The Josephson plasma mode. Is the SNS sandwich always a weak link?} \label{JPlasma}

Although the dynamical analysis is beyond the scope of the present work, we still want to address  the first elementary step of this  analysis:  the small oscillation around the ground state. For the  Josephson junction this is the Josephson plasma  oscillation. For an arbitrary  current--phase relation the Josephson plasma frequency is given by
\be
\omega_J = \sqrt{{2e\over C\hbar}{dJ(\theta)\over d\theta}},
    \ee{}
where $C$ is the capacitance of the Josephson junction and the derivative $dJ(\theta)/d\theta$ is  taken at $\theta$, which  corresponds to the ground state.
In usual Josephson junctions the Josephson plasma  frequency is much lower than the plasma  frequency
\be
\omega_0 =\sqrt{  4\pi e^2 n_0\over m}
  \ee{}
   in the bulk superconductor. This  inequality  is  in fact a necessary condition for  the existence of the Josephson plasma mode localized at the Josephson junction and decaying inside the superconducting  bulk.

Now let us consider a  3D  sandwich, which is a planar Josephson  SNS junction when the capacitance $C$ and the current $J(\theta)$ can be replaced by the capacity $4\pi/L$ per unit area
and the current density $J(\theta)/S$, where $S$ is the  area in the junction plane.  At zero temperature (more generally, at temperature much lower than the Andreev level energy spacing), the SNS sandwich near the ground state is in the regime of pure condensate charge transport, in which the vacuum and the excitation currents are absent, $\theta=\theta_s$, and according to \eq{vsJs},  $dJ(\theta)/d\theta=dJ_s(\theta_s)/d\theta_s= J_0/\pi$.  Then $\omega_J$ and $\omega_0$ coincide.  Thus, there is no localized Josephson plasma  mode. 

The localized Josephson plasma  mode in a Josephson junction  exists because  the  junction is a weak link. The hallmark of weak link is that the supercurrent through the junction requires  a phase gradient (ratio of the phase difference across  the junction to its length) much larger  than the phase gradient providing the same current in the bulk superconductor. The ballistic SNS sandwich at zero temperature is not a weak link in this meaning. 

\subsection{Meissner effect and Josephson vortices}  \label{Meissner}

Another manifestation that  due to the incommensurability effect the SNS sandwich is not always a weak link is its response to a weak magnetic field (Meisner effect). In the case of a usual planar Josephson junction the magnetic  field penetrates along the junction plane on  the Josephson penetration depth, which is much longer than the London penetration depth into the superconducting bulk. A planar ballistic SNS junction at zero temperature is not a weak link, and in the normal layer a supercurrent is supported  by the same phase gradient as in superconducting layers. Therefore,  the Josephson penetration depth does not differ from the London penetration depth.

Despite the SNS sandwich is not a weak link with respect to linear effects like the Josephson plasma oscillation or the Meisner effect, it is not the case  for nonlinear effects like the transition to the mixed state  at the first critical magnetic field.  
The first critical magnetic field is determined by the energy of the magnetic vortex localized near the normal layer (Josephson vortex).
Let us consider the Josephson vortex for the case when the London penetration depth $\lambda$ is much longer than the thickness $L$. So two inequalities are satisfied: $\lambda \gg L\gg \zeta_0$. The axis of the straight vortex is in the middle of the normal layer, and at distance $r$ from the axis exceeding $L$ the structure of the vortex does not differ essentially from the Abrikosov vortex in the superconductor bulk. The  area $r>L$ gives the  logarithmic  contribution to the vortex energy per vortex length:
\be
E_v =\left(\Phi_0\over 4\pi \lambda\right)^2\ln{\lambda\over L},
     \ee{Ev}
where $\Phi_0=hc/2e$ is the magnetic flux quantum. The area $r<L$ adds  a number of order unity  to the large logarithm.  The energy   $E_v$   is lower than the energy  of the Abrikosov vortex with the coherence length $\zeta_0$ replacing $L$ as a lower cut-off of the logarithm \cite{Tin}. 
If $L\gg \lambda$ the vortex energy is even smaller since the large logarithm in \eq{Ev} is replaced by a number of order unity.  This means that Josephson vortices are pinned to the normal layer, where their energy is less than the energy of Abrikosov  vortices in the superconducting layers. The vortex energy determines the first critical magnetic field: $H_{c1}=4\pi E_v/\Phi_0$.

\section{Summary and discussion} \label{Sum}

Previous investigations of the ballistic SNS sandwich were revised on the basis of our  approach, which properly satisfies the charge conservation law and takes  into account  the incommensurability of the superconducting gap with the Andreev level energy spacing.
Let us summarize the main conclusions of this work:
\begin{itemize}
\item
Due to the effect of incommensurability, in the ground state of a 1D ballistic SNS sandwich the phase difference $\theta$ across the sandwich is not necessarily 0, but  can take any value between 0 and $\pi$. Such a sandwich can be called $\theta$ junction. The  well known $\pi$ junction is a particular case $\theta=\pi$ of $\theta$ junctions.
\item
In 1D systems there is no essential suppression of the supercurrent through the ballistic SNS junction at temperatures on the order or  higher  than the energy distance between Andreev levels, but lower than the superconducting gap.
\item
Although the ballistic SNS junction has some properties of the Josephson  junction, it is not always a weak link in a strict sense. At zero temperature, or temperatures much lower than the Andreev level energy spacing, the weak magnetic field penetrates into the normal layer on the same London penetration depth as into the superconducting layers. There is no Josephson plasma mode localized at the normal layer in this case.
\item
The structure of magnetic vortices in the  ballistic SNS junction  essentially differs from structure of usual Josephson vortices, but still have energy lower than the energy of the Abrikosov vortex in the bulk of the superconductor. Therefore, vortices are pinned to the normal layer, and the first critical magnetic field for them is lower than for the superconductor bulk.

\end{itemize}

Through the whole paper the ballistic SNS sandwich  was considered as a Josephson junction. However, it is also possible to describe it not in terms of the Josephson physics. The ballistic normal  layer does not destroy the phase coherence and supports the supercurrent $J_s=en_0v_s$ with the same superfluid velocity $v_s$ and the same density $n_0$ as in the superconducting layers. The supercurrent is restricted by  the Landau criterion that the velocity $v_s$ does not exceeds 
 the Landau critical velocity equal at zero temperature to
\be
v_L = {\varepsilon_0\over \hbar k_f}={\pi \hbar \over 2mL}. 
     \ee{}
At this velocity the energy of a quasiparticle at the lowest Andreev level 
 becomes negative due to the Doppler shift.   
 This yields  the critical current $J_c=J_0$ given by \eq{kfn}. Since the Landau critical velocity inversely proportional to the layer thickness $L$, in the macroscopic (thermodynamic) limit $L\to \infty$ the Landau critical velocity vanishes.  Thus, ``superconductivity'' of the normal layer in the SNS sandwich is not a macroscopic, but a mesoscopic quantum phenomenon. 
 It is similar to  mesoscopic persistent currents in 1D normal metal rings predicted theoretically \cite{Buttiker,Gefen} and observed experimentally (see Ref.~\onlinecite{Moler} and references therein). The values of these persistent currents are  of the same order $ev_f/L$ as supercurrents in the ballistic SNS sandwich (for currents in normal rings $L$ is the circumference length of a ring). The origin of persistent currents was connected with  discreetness  of  energy levels in mesoscopic rings, but incommensurability is also an inevitable consequence of spectrum discreetness. In the case  of normal rings this is incommensurability of the Fermi energy (chemical potential) \cite{Gefen}, when the number of electrons changes from even to odd value. 
 In the case  of SNS sandwiches  the number of Andreev levels changes from even to odd. 

Analogy with persistent currents in mesoscopic normal ring points out a possible method of experimental investigation  of supercurrents in ballistic SNS sandwiches. In  normal rings they measured a magnetic moment induced by persistent currents as a function of the magnetic flux threading the ring. One can put  the SNS sandwich into a closed electrical circuit loop and make similar measurements. In fact, this idea has already been realized in the experiment on a carbon nanotube junction \cite{CNT}.   A carbon nanotube is a 1D, or, more accurately,  nearly  a 1D object (small number of active channels). \citet{CNT} observed the transition from 0 to $\pi$ junction in qualitative agreement with our prediction for 1D SNS junctions. Moreover, in the course of this transition they observed a current--phase relation with the period $\pi$ [see Fig.~4(b)(4) in their paper] two times smaller than the usual period $2\pi$. This is also expected from our analysis [see the paragraph after \eq{EJ}]. 

\citet{CNT} interpreted their experiment differently. They  considered a nanotube as a quantum dot and connected the 0--$\pi$ transition with the Kondo effect. Treating a nanotube as a quantum dot means that the  nanotube is rather short and the number of Andreev levels in it is not large. Our analysis is valid in the opposite limit of very long nanotube with large number of Andreev levels.  The fact that the 0--$\pi$ transition  is also  predicted in this limit means that the phenomenon is robust and not necessarily connected with the properties of quantum dots and the Kondo effect.

\begin{acknowledgments}
This work was started during my visit to the Low Temperature Laboratory of the Aalto University (Finland)  in October 2019. I thank Dmitry Golubev and Pertti Hakonen for numerous discussions and comments, which stimulated my interest to this problem and helped  its solution. I am also thankful  to the anonymous referee who detected a wrong sign of the vacuum current in the original version of this paper. This led to revision of some its conclusions. 
\end{acknowledgments}


%

\end{document}